\documentclass[conference]{IEEEtran}
\usepackage[noadjust]{cite}

\ifCLASSINFOpdf
  \usepackage[pdftex]{graphicx}
  \graphicspath{{./pdf/}}
  \DeclareGraphicsExtensions{.pdf,.jpeg,.png}
\else
  \usepackage[dvips]{graphicx}
  \graphicspath{{./eps/}}
  \DeclareGraphicsExtensions{.eps}
\fi
\usepackage{amsmath}
\usepackage{algorithmic}

\usepackage{array}
\usepackage{dblfloatfix}

\usepackage{url}

\usepackage{amssymb}    	
\usepackage{amsthm}      	
\usepackage[mathscr]{eucal}	
\usepackage[hidelinks]{hyperref}
\usepackage{enumerate}   	
\usepackage{dsfont}     	

\theoremstyle{plain}
\newtheorem{theorem}{Theorem}
\newtheorem{lemma}{Lemma}

\theoremstyle{remark}

\newtheorem*{remark*}{Remark}
\theoremstyle{definition}

\newtheorem{definition}{Definition}

\def\DD{\mathbb{D}}
\def\NN{\mathbb{N}}
\def\RR{\mathbb{R}}

\def\bge{\stackrel{\text{(b)}}{\ge}}

\def\ale{\stackrel{\text{(a)}}{\le}}
\def\ble{\stackrel{\text{(b)}}{\le}}

\def\aeq{\stackrel{\text{(a)}}{=}}

\def\ceq{\stackrel{\text{(c)}}{=}}
\def\deq{\stackrel{\text{(d)}}{=}}
\DeclareMathOperator{\var}{var}
\DeclareMathOperator{\ind}{\mathds{1}}	
\DeclareMathOperator{\E}{\mathbb{E}}
\DeclareMathOperator{\enc}{\mathscr{E}}
\DeclareMathOperator{\dist}{d_H}
\DeclareMathOperator{\weight}{w_H}
\def\cA{{\mathcal A}}
\def\cB{{\mathcal B}}
\def\cC{{\mathcal C}}
\def\cD{{\mathcal D}}
\def\cF{{\mathcal F}}
\def\cN{{\mathcal N}}
\def\cS{{\mathcal S}}
\def\cU{{\mathcal U}}
\def\cV{{\mathcal V}}
\def\indexset#1{{[\![#1]\!]}}

\def\typen{{\mathsf{N}}}	

\def\abs#1{\lvert#1\rvert}
\begin{document}
%
\title{On Channel Resolvability in Presence of Feedback}

\author{%
  \IEEEauthorblockN{Mani~Bastani~Parizi and Emre~Telatar}
  \IEEEauthorblockA{Information Theory Laboratory (LTHI), EPFL, Lausanne,
  Switzerland\\\{mani.bastaniparizi,emre.telatar\}@epfl.ch}
}


%


\maketitle

\begin{abstract}
  We study the problem of generating an approximately i.i.d.\ string at the
  output of a discrete memoryless channel using a limited amount of
  randomness at its input in presence of causal noiseless feedback.  Feedback
  does not decrease the \emph{channel resolution}, the minimum entropy rate
  required to achieve an accurate approximation of an i.i.d.\ output
  string. However, we show that, at least over a binary
  symmetric channel, a significantly larger
  \emph{resolvability exponent} (the exponential decay rate of the
  divergence between the output distribution and product measure), compared
  to the best known achievable resolvability exponent in a system without
  feedback, is possible.
  We show that by employing a variable-length resolvability scheme and
  using an average number of $R$ coin-flips per channel use, the
  \emph{average divergence} between the distribution of the output sequence
  and product measure decays exponentially fast in the \emph{average length
  of output sequence} with an exponent equal to $[R-I(U;V)]^+$ where
  $I(U;V)$ is the mutual information developed across the channel.
\end{abstract}


%
\IEEEpeerreviewmaketitle
\section{Introduction}
Suppose $P_{V|U}:\cU\to\cV$ is a discrete memoryless channel, with input
alphabet $\cU$ and output alphabet $\cV$, and we wish to
generate an i.i.d.\ string $V_1,V_2,\dotsc$ distributed according to $P_V$
at its output.   The obvious solution is to use an i.i.d.\ string
$U_1,U_2,\dotsc$ drawn from some distribution $P_U$, that induces $P_V$ at
the output of the channel, at its input which requires an entropy rate of
$H(U)$ bits per channel use (and results in a perfect i.i.d.\ output
sequence).  However, Wyner \cite{wyner:1975b} observed that, if we accept
an \emph{approximately i.i.d.\ sequence}, a lower entropy rate of $I(U;V)$
bits per channel use is sufficient (and necessary). Indeed, he showed that
if a random code of block-length $n$ and rate $R>I(U;V)$ is sampled from
i.i.d.\ $P_U$ random coding ensemble from which a uniformly chosen codeword
is transmitted via $n$ independent uses of the channel, with very high
probability over the choice of the code, the normalized Kullback--Leibler
divergence between the output distribution $P_{V^n}$ and the product
distribution $P_V^n(v^n) = \prod_{i=1}^n P_V(v_i)$,
$\frac1nD(P_{V^n}\|P_V^n)$ can be made arbitrarily small by choosing $n$
sufficiently large.  The problem of \emph{channel resolvability} was later
studied by Han and Verd\'u \cite{han:1993} and Hayashi \cite{hayashi:2006},
replacing the measure of approximation quality with total variation and
unnormalized divergence, respectively.
\begin{definition} \label{def:blockrate}
  A rate $R$ is \emph{achievable} over the channel $P_{V|U}:\cU\to\cV$ and
  with respect to (w.r.t.) the reference measure $P_V$ if there exists a
  sequence of $(n,k)$ codes, i.e., deterministic encoding functions
  $\enc^n: \{0,1\}^k\to\cU^n$, of rate at most $R$, 
  \begin{equation*}
    \limsup_{n\to\infty} \frac{k}{n} \le R,
  \end{equation*}
  such that, with $U^n=\enc^n(W^k)$, $W^k$
  uniformly distributed on $\{0,1\}^{k}$, and $V^n$ being the
  output of $n$ independent uses of $P_{V|U}$ with input $U^n$, denoted
  hereafter as $P_{V|U}^n$,
  \begin{equation} \lim_{n\to\infty}
    D(P_{V^n}\|P_V^n) = 0.
  \end{equation}
\end{definition}
\begin{definition} \label{def:resolution}
  The minimum of all achievable resolvability rates over the channel
  $P_{V|U}$ w.r.t.\ the reference measure $P_V$ is called the
  \emph{resolution} of the channel $P_{V|U}$ (w.r.t.\ to $P_V$).
\end{definition}
\begin{figure}
  \centerline{\includegraphics{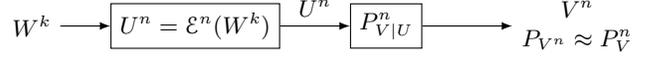}}
  \caption{Channel Resolvability}
  \label{fig:res}
\end{figure}
\begin{theorem}[{\cite{wyner:1975b,han:1993,hayashi:2006,hou:2013}}]
  \label{thm:blockrate}
  The resolution of the channel $P_{V|U}:\cU\to\cV$ w.r.t.\ the reference
  measure $P_V$ equals:
  \begin{equation}\label{eq:res}
   \min_{P_U: \sum_{u}P_U(u)P_{V|U}(v|u) = P_V(v)} I(U;V).
  \end{equation}
\end{theorem}
Moreover, in
\cite{hayashi:2006,hou:2013,han:2014,hayashi:2015arxiv,bastani:2015} it has
been shown that, in the above-mentioned context, the divergence between the
distribution of a length-$n$ block of channel output sequence $P_{V^n}$ and
product distribution $P_V^n$ decays exponentially fast in $n$ 
and in \cite{bastani:2016} the exact exponential decay rate of
the ensemble-average of $D(P_{V^n}\|P_V^n)$ as a function of $R$ is
characterized.
\begin{definition} \label{def:blockexp}
  A pair $(R,E)$ is an \emph{achievable} resolvability rate--exponent pair
  over the channel $P_{V|U}:\cU\to\cV$ w.r.t.\ the reference measure $P_V$
  if there exists a sequence of $(n,k)$ codes  
  $\enc^n:\{0,1\}^k\to\cU^n$ of rate at most $R$,
  \begin{equation*}
    \limsup_{n\to\infty} \frac{k}{n} \le R,
  \end{equation*}
  such that, with $U^n=\enc^n(W^k)$, $W^k$ uniformly distributed over
  $\{0,1\}^k$, and $V^n$ being the output of $P_{V|U}^n$ to input $U^n$,
  \begin{equation}
    \liminf_{n\to\infty} -\frac1n \log D(P_{V^n}\|P_V^n) \ge E.
  \end{equation}
\end{definition}
\begin{theorem}[\cite{bastani:2016}]
  Suppose the encoder in Fig.~\ref{fig:res} is a code of rate $R$
  constructed randomly by sampling from i.i.d.\ $P_U$ random coding
  ensemble, $\{U^n(w^k)\colon w^k\in\{0,1\}^k\}$, $k = \lfloor nR \rfloor$,
  and outputs $\enc^n(w^k) = U^n(w^k)$.  Then (when $W^k$ is uniformly
  distributed on $\{0,1\}^k$),
  \begin{multline} \label{eq:blockexp}
    \lim_{n\to\infty} -\frac1n
    \log\bigl(\overline{D(P_{V^n}\|P_V^n)}\bigr)\\
    = \min_{Q_{UV}}\{D(Q_{UV}\|P_{UV})+[R-f(Q_{UV}\|P_{UV})]^+\},
  \end{multline}
  where, $\overline{D(P_{V^n}\|P_V^n)}$ is the ensemble-average of
  $D(P_{V^n}\|P_V^n)$, 
  $$f(Q_{UV}\|P_{UV})\triangleq \sum_{u,v} Q_{UV}(u,v)
  \log\frac{P_{V|U}(v|u)}{P_{V}(v)},$$
  and $P_V(v)=\sum_{u} P_U(u)P_{V|U}(v|u)$.
\end{theorem}
\begin{remark*}
  The achievability of the exponent \eqref{eq:blockexp} was shown in
  \cite{han:2014,hayashi:2015arxiv,bastani:2015} and its
  exactness is established in \cite{bastani:2016}.  To the extent of our
  knowledge, the exponent of
  \eqref{eq:blockexp} is the best achievable resolvability exponent
  reported so far in the literature.
\end{remark*}
\begin{figure}
  \centerline{\includegraphics{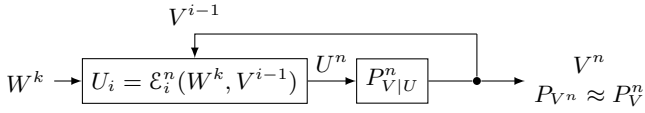}}
  \caption{Channel Resolvability in Presence of Feedback}
  \label{fig:feedback}
\end{figure}
In this paper we consider the problem of channel resolvability in presence
of causal feedback,  namely, when the encoder gets to know the past
received symbols $V^{i-1}$ before transmitting the $i$\textsuperscript{th}
symbol $U_i$ and, hence, have the opportunity of deciding about the value
of $U_i$ based on the past behavior of the channel (see
Fig.~\ref{fig:feedback}).  

Channel resolvability is, in a sense, the countrpart of channel coding.
For channel coding, it is well-known that feedback does not increase the
channel capacity \cite[Exercise~4.6]{gallager:it}.  Likewise, feedback does
not reduce the channel resolution (see Theorem~\ref{thm:rate}).  On
the other hand, Burnashev \cite{burnashev:1976} showed that, in presence of
feedback (and using variable-length codes) higher error exponents are
achievable.  Thus, it is natural to ponder if the same holds for channel
resolvability?

In this work, we give an affirmative answer to the above, at least when the
channel $P_{V|U}:\cU\to\cV$ is a binary symmetric channel (BSC) and the
reference measure $P_V$ is uniform on $\{0,1\}$. We show that in presence
of causal feedback and using variable-length resolvability codes the
straight-line exponent $[R-I(U;V)]^+$ is achievable (see
Theorem~\ref{thm:feedbackexp}).  
\section{Preliminaries}
\subsection{Notation}
We use uppercase letters (like $U$) to denote a random variable and
the corresponding lowercase version ($u$) for a realization of that random
variable.  The same convention applies to the sequences, i.e., $u^n =
(u_1,\dotsc,u_n)$ denotes a realization of the random sequence
$U^n=(U_1,\dotsc,U_n)$. If $\cS$ is a finite set, $|\cS|$ denotes its
cardinality.  Given an alphabet $\cA$, $\cA^\ast$ denotes the set
of all strings over symbols in $\cA$.
Given a pair of real numbers $a<b$, $\indexset{a:b}\triangleq[a,b]\cap\NN$
denotes the set of integers between $a$ and $b$.  For $a \in \RR$,
$[a]^+\triangleq\max\{a,0\}$.

Binary divergence $d_2(\cdot\|\cdot)$, binary entropy function
$h_2(\cdot)$, and binary capacity function $c_2(\cdot)$ are defined,
respectively as
\begin{align}
  d_2(p\|q) &\triangleq p\log\frac{p}{q} + (1-p)\log\frac{1-p}{1-q},\\
  h_2(p) & \triangleq p\log\frac{1}{p} + (1-p)\log\frac{1}{1-p}, \quad
  \text{and}\\
  c_2(p) & \triangleq 1-h_2(p). \label{eq:c2def}
\end{align}
Finally, $\weight(u^n)$ denotes the Hamming weight of the binary sequence
$u^n$ and $\dist(u^n,v^n)=\weight(u^n\oplus v^n)$ denotes the Hamming
distance between two sequences $u^n$ and $v^n$.
\subsection{Resolvability with Variable-Length Codes}
The classical channel resolvability problem is defined based on block
codes.  Namely, the aim is to make the distribution of a length-$n$ block of
the output $P_{V^n}$ close to i.i.d.\ $P_V^n$ using a $(n,k)$ block code of
rate at most $R$ and $k \le nR$ coin-flips at the encoder. It is useful
extend this notion to variable-length codes.  Specifically, the encoder is
confined to use only $k$ coin-flips but is allowed to use the channel a
variable number
of times based on a stopping rule.  
\begin{definition}\label{def:varcode}
  A \emph{$(\ast,k)$ variable-length resolvability code} (or in short a
  $(\ast,k)$ resolvability code), in presence of feedback, over the input
  and output alphabets $(\cU,\cV)$ is defined via a collection of
  deterministic encoding functions 
  \begin{equation} \label{eq:enc}
    \enc^{(k)}_n\colon \{0,1\}^k \times \cV^{n-1} \to \cU \cup \{\mathsf{S}\},
    \qquad n \in \NN,
  \end{equation}
  where $\mathsf{S}\not\in\cU$ is a special symbol indicating the ``end of
  transmission.''  Namely, given the input word $w^k$ and the past channel
  output symbols $V^{n-1}$,  the encoding function $\enc^{(k)}_n$ decides
  to either feed the channel with an input symbol in $\cU$ or stop the
  encoding (by outputting $\mathsf{S}$).

  Given a {$(\ast,k)$ resolvability code}, a \emph{$(\ast,k)$ feedback
  resolvability encoder} maps the input word $w^k$ into a
  channel input sequence $U_1,U_2,\dotsc$ as follows:
  \begin{algorithmic}[1]
    \STATE $n\gets1$;
    \WHILE {$\enc_n^{(k)}(w^k,V^{n-1}) \ne \mathsf{S}$}
    \STATE $U_n \gets \enc_n^{(k)}(w^k,V^{n-1})$;
    \STATE Transmit $U_n$ via the channel $P_{V|U}:\cU\to\cV$;
    \STATE $n\gets n+1$;
    \ENDWHILE
  \end{algorithmic}
\end{definition}
\begin{remark*}
  A $(n,k)$ block resolvability code is a special case of
  a $(\ast,k)$ variable-length resolvability code.
\end{remark*}
Obviously, when a variable-length feedback resolvability encoder is
employed, the \emph{stopping time} of the encoder (and hence the length of
the channel output corresponding to a single run of the encoder) will be a
random variable, which we denote by $N_k$, that depends both on the channel
randomness and the randomness of the input word $W^k$. 
We  measure the performance of the system by the expected output divergence
\begin{equation} \label{eq:divdef}
  \DD_k \triangleq \sum_{n} D(P_{V^n|N_k=n}\|P_V^n) \Pr\{N_k=n\}
\end{equation} 
and the expected number of channel uses, $\E[N_k]$. Indeed, by the law of
large numbers, when the resolvability scheme is run a large number of
times (each corresponding to a block of channel output), the output
sequence will have an average length of $\E[N_k]$ symbols per block and the
divergence between distribution of the output string and the product
distribution normalized by the number of blocks will be close to
$\DD_k$. We can, hence, extend Definitions~\ref{def:blockrate} and
\ref{def:blockexp} as:
\begin{definition} \label{def:rate}
  $R$ is an \emph{achievable} resolvability rate over the channel
  $P_{V|U}:\cU\to\cV$ w.r.t.\ the reference measure $P_V$ if there exists a
  sequence of $(\ast,k)$ resolvability codes (cf.\
  Definition~\ref{def:varcode}) such that, when $W^k$ is uniformly
  distributed on $\{0,1\}^k$,
  \begin{equation}
    \limsup_{k\to\infty} \frac{k}{\E[N_k]} \le R,
  \end{equation}
  and, with $\DD_k$ defined as in \eqref{eq:divdef},
  \begin{equation} \label{eq:strongres}
    \lim_{k\to\infty} \DD_k = 0.
  \end{equation}
\end{definition}
\begin{definition} \label{def:exp}
  A pair $(R,E)$ is an \emph{achievable} resolvability rate--exponent pair
  over the channel $P_{V|U}:\cU\to\cV$ w.r.t.\ the reference measure $P_V$
  if there exists a sequence of $(\ast,k)$
  resolvability codes (see Definition~\ref{def:varcode}) such that, when
  $W^k$ is uniformly distributed on $\{0,1\}^k$,
  \begin{equation}
    \limsup_{k\to\infty} \frac{k}{\E[N_k]} \le R,
  \end{equation}
  and, with $\DD_k$ defined as in \eqref{eq:divdef},
  \begin{equation}
    \liminf_{k\to\infty} - \frac{\log \DD_k}{\E[N_k]} \ge E.
  \end{equation}
\end{definition}
\section{Results}
\begin{figure}
  \centerline{\includegraphics{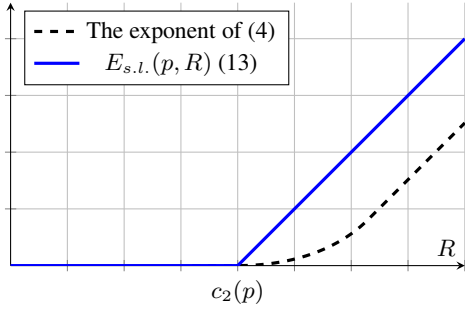}}
  \caption{Comparison of the exponents}
  \label{fig:comp}
\end{figure}

\begin{theorem}\label{thm:rate}
  Employing variable-length resolvability codes (in presence of
  feedback) does not reduce the channel resolution.
\end{theorem}
\begin{theorem} \label{thm:feedbackexp}
  In presence of feedback, the exponent
  \begin{equation} \label{eq:feedbackexp} 
    E_{s.l.}(p,R) = [R-c_2(p)]^+
  \end{equation}
  is achievable via a sequence of variable-length resolvability codes over
  a BSC with crossover probability $p$ with respect to the uniform
  reference measure $P_V(0)=P_V(1)=\frac12$.
\end{theorem}
\begin{remark*} 
  The straight-line exponent of \eqref{eq:feedbackexp} is larger than the
  exponent
  of \eqref{eq:blockexp} as the objective function of \eqref{eq:blockexp}
  equals $[R-I(U;V)]^+$ at $Q_{UV}=P_{UV}$ (see Fig.~\ref{fig:comp}).
\end{remark*}
\section{Proofs}
\subsection{Proof of Theorem~\ref{thm:rate}}
We prove the converse under \emph{weak} resolvable criteria which implies
that under strong resolvability criteria, \eqref{eq:strongres}. Accordingly,
assume we have a sequence of $(k,\ast)$ codes satisfying
\begin{equation}\label{eq:gegenteil:weakres}
  \limsup_{k\to\infty} \frac{\DD_k}{\E[N_k]} = 0.
\end{equation}

Let $U^\infty$ and $V^\infty$ denote the infinite channel input and output
sequences with $U_m=\mathsf{S}$ and $V_m=\varnothing\notin\cV$ if the
transmission stops before time $m$. Let also $\chi_m\triangleq\ind\{N_k\ge
m\}$. Therefore,
\begin{align}
  & k=H(W^k)\ge I(W^k,V^\infty)
  = \sum_{m\geq 1} I(W^k,V_m|V^{m-1})\nonumber\\
  &\quad = \sum_{m\geq 1} [H(V_m|V^{m-1})-H(V_m|W^k,V^{m-1})]\nonumber\\ 
  &\quad \aeq \sum_{m\geq 1} [H(V_m|V^{m-1})-H(V_m|W^k,V^{m-1},U_m,\chi_m)]
  \nonumber \\
  &\quad \bge \sum_{m\geq 1} [H(V_m|V^{m-1},\chi_m)-H(V_m|U_m,\chi_m)].
  \label{eq:gegenteil:sumdiff}
\end{align}
In the above, (a) follows since $U_m=\enc_m^{(k)}(W^k,V^{m-1})$, and
$\chi_m=\ind\{U_m\ne\mathsf{S}\}$ according to
Definition~\ref{def:varcode} and (b) since conditioning reduces the
entropy.
Now, observe that
\begin{equation*}
  H(V_m|V^{m-1},\chi_m)=H(V_m|V^{m-1},N_k\ge m)\Pr\{N_k\ge m\}
\end{equation*}
since $\{N_k<m\}$ implies $V_m=\varnothing$.   Let
\begin{equation}
  \beta(\delta)\triangleq
  \sqrt{2\ln(2)\delta}\log\frac{\abs{\cV}}{\sqrt{2\ln(2)\delta}}.
\end{equation}
The uniform continuity of entropy
\cite[Lemma~2.7]{csiszar:it} together with Pinsker's inequality
and Jensen's inequality imply
\begin{multline}
  \abs{H(V_m|V^{m-1},N_k\ge m)-H(V)}\\
  \le \beta\bigl(D(P_{V_m|V^{m-1},N_k\ge m}\|P_V|P_{V^{m-1}|N_k\ge
  m})\bigr)
\end{multline}
Consequently,
\begin{align}
  &\sum_{m\geq 1}H(V_m|V^{m-1},\chi_m) \nonumber\\
  & \quad = \sum_{m\geq 1} H(V_m|V^{m-1},N_k\geq m) \Pr\{N_k\geq m\}
  \nonumber\\
  & \quad \ge H(V) \sum_{m\geq 1}\Pr\{N_k\geq m\} - \sum_{m\geq 1}\bigl[\Pr\{N_k\geq
  m\} \nonumber \\ & \quad\qquad \cdot  \beta\bigl(D(P_{V_m|V^{m-1},N_k\ge
  m}\|P_V|P_{V^{m-1}|N_k\ge m})\bigr)\bigr] \nonumber\\
  & \quad = H(V) \E[N_k] - \sum_{m\geq 1}\bigl[\Pr\{N_k\geq
  m\} \nonumber \\ & \quad\qquad \cdot  \beta\bigl(D(P_{V_m|V^{m-1},N_k\ge
  m}\|P_V|P_{V^{m-1}|N_k\ge m})\bigr)\bigr] \nonumber\\
  & \quad \stackrel{(\ast)}{\geq} \E[N_k] \Biggl[H(V)-\beta\Biggl(\sum_{m\geq 1}
      \frac{\Pr\{N_k\geq m\}}{\E[N_k]} \nonumber \\
  & \quad \qquad \cdot D(P_{V_m|V^{m-1},N_k\geq m}\|P_V|P_{V^{m-1}|N_k\geq
  m})\Biggr)\Biggr],
  \label{eq:gegenteil:entv}
\end{align}
where $(\ast)$ follows by concavity of $\beta$. On the other hand, the
convexity of divergence implies
\begin{multline*}
  \Pr\{N_k\ge m\}D(P_{V_m|V^{m-1},N_k\ge m}\|P_V|P_{V^{m-1}|N_k\geq m}) \\
  \le \sum_{n\ge m} D(P_{V_m|V^{m-1},N_k=n}\|P_V|P_{V^{m-1}|N_k=n})
  \Pr\{N_k=n\}.
\end{multline*}
Therefore,
\begin{align}
  & \sum_{m\geq 1} \Pr\{N_k\ge m\}D(P_{V_m|V^{m-1},N_k\ge
  m}\|P_V|P_{V^{m-1}|N_k\geq m}) \nonumber \\
  & \enspace \le \sum_{\substack{m\geq 1,\\n\geq m}}
  D(P_{V_m|V^{m-1},N_k=n}\|P_V|P_{V^{m-1}|N_k=n})
  \Pr\{N_k=n\}\nonumber\\
  & \enspace = 
  \sum_{n\geq 1} \Pr\{N_k=n\} \nonumber \\
  & \enspace \phantom{=\sum_{n\geq 1}}\qquad \cdot \sum_{m=1}^{n} 
  D(P_{V_m|V^{m-1},N_k=n}\|P_V|P_{V^{m-1}|N_k=n})\nonumber\\
  & \enspace \stackrel{(\ast)}{=}
  \sum_{n\geq 1}D(P_{V^n|N_k=n}\|P_V^n)\Pr\{N_k=n\},
  \label{eq:gegenteil:divavg1}
\end{align}
where $(\ast)$ follows by the chain rule.  Using
\eqref{eq:gegenteil:divavg1} in \eqref{eq:gegenteil:entv} together with the
fact that $\beta$ is an increasing function, we conclude that
\begin{equation}
  \sum_{m\ge 1}H(V_m|V^{m-1},\chi_m) \ge \E[N_k] \Bigl[H(V) -
  \beta\Bigl(\frac{\DD_k}{\E[N_k]}\Bigr)\Bigr]
  \label{eq:gegenteil:entv2}
\end{equation}

Similarly, we have
\begin{equation*}
  H(V_m|U_m,\chi_m)\\ = H(V_m|U_m,N_k\ge m)\Pr\{N_k\ge m\}.
\end{equation*}
Now note that 
$P_{V_m|U_m,N_k\ge m}(v|u)=P_{V|U}(v|u)$, therefore, defining
\begin{equation}
  \gamma(\delta) \triangleq \max_{P_U: D(P_U\circ P_{V|U}\|P_V)\le \delta}
  H(V|U)
\end{equation}
(where we have used the shorthand notation $(\!P_U\circ
P_{V|U}\!)(v)\triangleq\sum_{u}P_U(u)P_{V|U}(v|u)$),
\begin{equation}
  H(V_m|U_m,N_k\ge m) \le  \gamma\bigl(D(P_{V_m|N_k\ge m}\|P_V)\bigr).
\end{equation}
Noting that $\gamma$ is
concave%
\footnote{It can be verified that if $f(x)\colon\cD\to\RR$ is convex and
  $l(x)\colon\cD\to\RR$ is a linear function of $x$, (on some convex domain
  $\cD$) then the mapping $y \mapsto \max_{x:f(x)\le y}l(x)$ is concave in
  $y$.
},%
similar steps as \eqref{eq:gegenteil:entv} yield
\begin{multline}
  \sum_{m\geq 1}H(V_m|U_m,\chi_m) \\ \le \E[N_k] \gamma\Biggl(\sum_{m\geq
  1}\frac{\Pr\{N_k\geq m\}}{\E[N_k]} D(P_{V_m|N_k\geq m}\|P_V)\Biggr).
  \label{eq:gegenteil:condent1}
\end{multline}
Once again, the convexity of divergence implies
\begin{multline*}
  \Pr\{N_k\ge m\}D(P_{V_m|N_k\ge m}\|P_V) \\
  \le \sum_{n\ge m} D(P_{V_m|N_k=n}\|P_V)
  \Pr\{N_k=n\},
\end{multline*}
and same steps as \eqref{eq:gegenteil:divavg1} show
\begin{multline}
  \sum_{m\geq1}\Pr\{N_k\geq m\}D(P_{V_m|N_k\geq m}\|P_V) \\
  \le \sum_{n\geq 1}\Biggl(\sum_{m=1}^{n}D(P_{V_m|N_k=n}\|P_V)\Biggr) 
  \Pr\{N_k=n\}
  \label{eq:gegenteil:divavg2a}
\end{multline}
Since 
\begin{multline*}
  D(P_{V^n|N_k=n}\|P_V^n) 
  = D\Bigl(P_{V^n|N_k=n}\Big\|\prod_{m=1}^{n} P_{V_m|N_k=n}\Bigr) 
  \\
  + \sum_{m=1}^{n} D(P_{V_m|N_k=n}\|P_V),
\end{multline*}
we can further upper-bound the term inside the parenthesis in
\eqref{eq:gegenteil:divavg2a} by $D(P_{V^n|N_k=n}\|P_V^n)$ to conclude that
\begin{equation}
  \sum_{m\geq1}\Pr\{N_k\geq m\}D(P_{V_m|N_k\geq m}\|P_V) 
  \le \DD_k.  \label{eq:gegenteil:divavg2}
\end{equation}
Using \eqref{eq:gegenteil:divavg2} and the fact that $\gamma$ is
increasing in \eqref{eq:gegenteil:condent1} we get
\begin{equation}
  \sum_{m\geq 1}H(V_m|U_m,\chi_m) \le \E[N_k]
  \gamma\Big(\frac{\DD_k}{\E[N_k]}\Bigr).
  \label{eq:gegenteil:condent}
\end{equation}

Finally, uniting \eqref{eq:gegenteil:entv2} and
\eqref{eq:gegenteil:condent} in \eqref{eq:gegenteil:sumdiff} yields
\begin{equation} \label{eq:gegenteil:last}
  \frac{k}{\E[N_k]} \ge H(V)-\gamma\Bigl(\frac{\DD_k}{\E[N_k]}\Bigr) -
  \beta\Bigl(\frac{\DD_k}{\E[N_k]}\Bigr).
\end{equation}
Since $\lim_{\delta\to0} \beta(\delta)=0$ and, as $H(V|U)$ is continuous in
$P_U$, $\lim_{\delta\to
  0}\gamma(\delta) = \max_{P_U: P_U\circ P_{V|U}=P_V} H(V|U)$, \eqref{eq:gegenteil:last} together with
the assumption \eqref{eq:gegenteil:weakres} yield
\begin{equation}
  \liminf_{k\to\infty} \frac{k}{\E[N_k]} \ge \min_{P_U: P_U\circ
  P_{V|U}=P_V} I(U;V).\tag*{\IEEEQED}
\end{equation}
\subsection{Proof of Theorem~\ref{thm:feedbackexp}}
To prove Theorem~\ref{thm:feedbackexp},
we propose the following sequence of $(\ast,k)$ resolvability codes and
show that the exponent of \eqref{eq:feedbackexp} is achievable using this
sequence of codes.  Throughout the proof, without essential loss of
generality, we assume $p < \frac12$.  

\paragraph*{Proposed Sequence of Codes} 
Fix $\alpha > 0$.  We define a $(\ast,k)$ code for each $k$ as follows:
The collection of encoding functions $(\enc^{(k)}_n,\,n\in\NN)$ share a
codebook of size $2^k$ and infinite block-length indexed by
length-$k$ binary sequences, $\cC_k\triangleq\{u^\infty(w^k): w^k\in\{0,1\}^k\}$ (to be
specified later) and are defined as
\begin{subequations}
\begin{align}
  \enc^{(k)}_1(w^k) &= u_1(w^k), \quad \text{and}\\
  \enc_{n+1}^{(k)}(w^k,V^n) &= \begin{cases}
    \mathsf{S} & \text{if $\frac{k}{n} \le \alpha
    c_2(\hat{Q}_n)$},\\
    u_{n+1}(w^k) & \text{otherwise,}
  \end{cases}
\end{align}
\end{subequations}
where $$\hat{Q}_n\triangleq\frac{\dist(u^n(w^k),V^n)}{n}$$ is the fraction
of flipped bits in the time interval of $\indexset{1:n}$.

Namely, given the input word $w^k$, the encoder transmits the corresponding
codeword $u^{\infty}(w^k)$ bit-by-bit until the transmission rate
$\frac{k}{n}$ drops below $\alpha$ times the empirical capacity of the
channel.  Consequently, the stopping $N_k$ is larger than
$\frac{k}{\alpha}$.
\begin{lemma} \label{lem:stop}
  For the proposed scheme,
  \begin{equation}
    \lim_{k\to\infty} \frac{k}{\E[N_k]} = \alpha c_2(p).
  \end{equation}
\end{lemma}
\begin{IEEEproof}
  Let $B_n\triangleq\ind\{\text{channel flips at time $n$}\}.$ Hence
  $n \hat{Q}_n=\sum_{j=1}^nB_j$ where
  $(B_n,\,n\in\NN)$ are i.i.d.\ $\mathrm{Bernoulli}(p)$  random
  variables.  Let $S_n\triangleq n\hat{Q}_n-np,$ and observe that the process
  $(S_n,\,n\in\NN)$ is a martingale w.r.t.\ the natural filtering
  $\bigl(\cF_n=\sigma(B_1,\dots,B_n),\,n\in\NN\bigr)$.  The encoder
  stops at time
  \begin{equation} \label{eq:stoptime}
    N_k=\inf\Bigl\{n\ge\frac{k}{\alpha}\colon
    c_2\bigl(\hat{Q}_n\bigr)\ge\alpha^{-1} \frac{k}{n}\Bigr\}.
  \end{equation}
  In terms of $S_n$ the stopping condition is
  \begin{equation} \label{eq:stop:upper1}
    k \le \alpha \cdot N_k c_2\Bigl(p+\frac{S_{N_k}}{N_k}\Bigr).
  \end{equation}
  It easily can be verified that $\forall p \in (0,1)$, $\forall
  \varepsilon \in
  (-p,1-p)$,
  \begin{equation}
    c_2(p) + \varepsilon c_2'(p) \le
    c_2(p+\varepsilon) \le c_2(p) + c_2'(p) \varepsilon + c_2''(p)
    \varepsilon^2
    \label{eq:c2parabola}
  \end{equation}
  Using the upper bound of \eqref{eq:c2parabola} in \eqref{eq:stop:upper1}
  we get
  \begin{equation} \label{eq:stop:upper2}
    k \le \alpha c_2(p)  N_k  + \alpha c_2'(p)  S_{N_k} + 
    \alpha c_2''(p) \frac{S_{N_k}^2}{N_k}.
  \end{equation}
  Taking the expectation of the right-hand-side of \eqref{eq:stop:upper2},
  noting that $\E[S_{N_k}]=\E[S_{\lceil k/\alpha\rceil}]=0$ (because
    a stopped martingale is also a martingale \cite[Theorem~4,
  Chapter~7]{gallager:dsp}), we get
  \begin{equation} \label{eq:stop:upper3}
    \frac{k}{\E[N_k]} \le \alpha c_2(p) + \alpha c_2''(p)
    \frac{\E[S_{N_k}^2/N_k]}{\E[N_k]}.
  \end{equation}
  It remains to examine the growth rate of the last term in
  \eqref{eq:stop:upper3}.  Had we replaced the stopping time $N_k$
  with a fixed time $n$, the quantity of interest would have behaved like
  $\frac1n$ (since $\E[S_n^2/n]$ is a constant).   
  It turns out that for a stopping time $N_k$, $\E[S_{N_k}^2/N_k]$ may not
  be a constant but will grow at most logarithmically in $N_k$:  
  Lemma~\ref{lem:stopvar} (in the appendix) shows
  \begin{equation}
    \E\Bigl[\frac{S_{N_k}^2}{N_k}\Bigr] \le p(1-p)\E[1+\ln(N_k)].
  \end{equation}
  Consequently,
  \begin{align}
    \frac{k}{\E[N_k]}
    & \le \alpha c_2(p) + \alpha c_2''(p) p (1-p)
    \frac{\E[1+\ln(N_k)]}{\E[N_k]} \nonumber\\
    &\ale \alpha c_2(p) + \alpha c_2''(p) p (1-p)
    \frac{1+\ln(\E[N_k])}{\E[N_k]} \nonumber\\
    &\ble \alpha c_2(p) + \alpha c_2''(p) p (1-p)
    \frac{1 + \ln(k/\alpha)}{k/\alpha},
    \label{eq:stop:upper4}
  \end{align}
  where (a) follows from Jensen's inequality and (b) as
  $\frac{1+\ln(x)}{x}$ is decreasing for $x\ge1$ and $N_k\ge
  \frac{k}{\alpha}$. Consequently,
  \begin{equation}
    \limsup_{k\to\infty}\frac{k}{\E[N_k]}\le \alpha c_2(p).
    \label{eq:stop:sup}
  \end{equation}

  To lower-bound $k/\E[N_k]$, we note that $\forall n>1$, $\hat{Q}_n =
  \frac{n-1}{n} \hat{Q}_{n-1} + \frac1n B_n$.  Since $c_2(\cdot)$ is
  convex, at the stopping time,
  \begin{multline}
    c_2(\hat{Q}_{N_k}) \le \frac{{N_k}-1}{N_k}c_2(\hat{Q}_{N_k-1}) 
    + \frac{1}{N_k} c_2(B_{N_k}) \\
    \stackrel{(\ast)}{<} \alpha^{-1} \frac{N_k-1}{N_k}
    \times \frac{k}{N_k-1} + \frac{1}{N_k} 
     = \alpha^{-1} \frac{k}{N_k} + \frac{1}{N_k}.
  \end{multline}
  where $(\ast)$ follows from the stopping condition \eqref{eq:stoptime}.
  Therefore, substituting $\hat{Q}_{N_k}=\frac{S_{N_k}}{N_k}+p$, 
  \begin{align}
    k & > \alpha N_k c_2\Bigl(p+\frac{S_{N_k}}{N_k}\Bigr)-\alpha  \nonumber
    \\ 
    & \ge \alpha c_2(p) N_k + \alpha c_2'(p) S_{N_k} - \alpha,
    \label{eq:stop:lower1}
  \end{align}
  where the second inequality \eqref{eq:stop:lower1} follows from the lower
  bound of \eqref{eq:c2parabola}.  Taking the expectation of the
  right-hand-side of \eqref{eq:stop:lower1} (and using the fact that
  $\E[S_{N_k}]=0$ once again) we get,
  \begin{equation} \label{eq:stop:lower2}
    \frac{k}{\E[N_k]} \ge \alpha c_2(p) - \frac{\alpha}{\E[N_k]} \ge \alpha
    c_2(p) - \frac{\alpha^2}{k}.
  \end{equation}
  where the second inequality follows since $N_k\ge k/\alpha$. 
  Thus,
  \begin{equation}\label{eq:stop:inf}
    \liminf_{k\to\infty} \frac{k}{\E[N_k]}
    \ge \alpha c_2(p),
  \end{equation}
  which, together with \eqref{eq:stop:sup} concludes the proof.
\end{IEEEproof}

To complete the proof of Theorem~\ref{thm:feedbackexp}, it remains to
bound the expected output divergence $\DD_k$ \eqref{eq:divdef} for an
appropriate code.   

\begin{figure}
  \centerline{\includegraphics{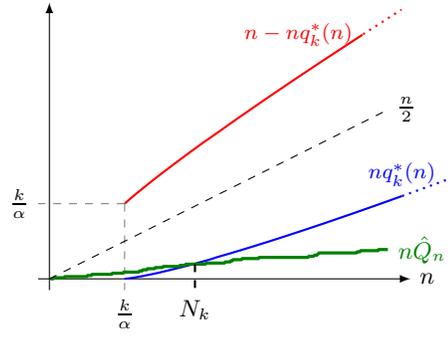}}
  \caption{Encoder's Stopping Time}
  \label{fig:stop}
\end{figure}

Let $c_2^{-1}(\cdot)$ denote the inverse of the binary capacity function
$c_2(\cdot)$ (cf.~\eqref{eq:c2def}) when its domain is restricted to
$[0,\frac12]$ and define $q_k^\star\colon\indexset{k/\alpha:+\infty}\to[0,\frac12]$:
\begin{equation} \label{eq:qstardef}
  q_k^*(n) \triangleq c_2^{-1}\Bigl(\alpha^{-1} \frac{k}{n}\Bigr).
\end{equation}
Let $B^n\triangleq(B_1,\dotsc,B_n)$ denote the flip pattern of $n$
independent uses of the channel and 
\begin{equation}
  \cB_n \triangleq \bigl\{b^n\in\{0,1\}^n: \{B^n=b^n\}\subset\{N_k=n\}
  \bigr\}
\end{equation}
denote the set of flip patterns that stop the encoder at time $N_k=n$.  Using
the fact that the process $n\hat{Q}_n = \weight(B^n)$ is an integer-valued
process and the stopping condition \eqref{eq:stoptime} we can conclude that
(among other constraints) $\forall b^n\in\cB_n$, either $\weight(b^n) =
\lfloor nq^*_k(n) \rfloor$ or $n-\weight(b^n)=\lfloor n q^*_k(n)\rfloor$
(see Fig.~\ref{fig:stop}).  

Note that $\cB_n$ can be empty for some values of $n\in\indexset{k/\alpha:+\infty}$.\footnote{%
  For example, if for some $n$, $\exists \ell\in\NN$ such that
  $\lfloor(n-\ell)q_k^*(n-\ell)\rfloor=\lfloor nq_k^*(n)\rfloor$ then,
  $\cB_n$ is empty because either the encoder stops at time $n-\ell$ or, if
  not, it will stop at some time $N_k>n$, because $$\weight(B^{n}) \ge
  \weight(B^{n-\ell}) > \lfloor n q_k^*(n) \rfloor$$ and similarly
  $n-\weight(B^{n}) < n-\lfloor n q_k^*(n) \rfloor.$}
Obviously for such $n$s $\Pr\{N_k=n\}=0$ so we shall not be concerned about
them.  Let $$\cN_k \triangleq \{n\in\indexset{k/\alpha:+\infty}:
\Pr\{N_k=n\} > 0\}$$ be the support of $N_k$ and assume $n\in\cN_k$.

Partition $\cB_n = \cB_n^1\cup\cB_n^2$ where
\begin{align*}
  \cB^1_n &\triangleq \{b^n\in\cB_n: \weight(b^n)=\lfloor n q_k^*(n)
  \rfloor\}, \\
  \cB^2_n &\triangleq \{b^n\in\cB_n: \weight(b^n)=n-\lfloor n q_k^*(n)
  \rfloor\}.
\end{align*} 
It can easily be verified that $|\cB^1_n|=|\cB^2_n|=\frac12|\cB_n|$.
Indeed, the symmetry of stopping thresholds around $\frac{n}{2}$
(Fig.~\ref{fig:stop}) implies $b^n\in\cB^1_n$ if and only if $b^n \oplus
\mathbf{1}^n \in \cB^2_n$ (where $\mathbf{1}^n$ denotes the all-one vector
of length $n$).  Consequently,
\begin{subequations}\label{eq:probb}
  \begin{align}
    \Pr\{B^n\in\cB_n^1\} &= \frac12|\cB_n| p^{\lfloor nq_k^*(n)\rfloor}
    (1-p)^{n-\lfloor n q_k^*(n) \rfloor},\\
    \Pr\{B^n\in\cB_n^2\} &= \frac12|\cB_n| p^{n-\lfloor nq_k^*(n)\rfloor}
    (1-p)^{\lfloor n q_k^*(n) \rfloor}.
  \end{align}
\end{subequations}
Since $0\le p \le \frac12$,
$\Pr\{B^n\in\cB_n^1\}\ge\Pr\{B^n\in\cB_n^2\}$.  Hence,
$$\rho_n\triangleq\frac{\Pr\{B^n\in\cB_n^1\}}{\Pr\{B^n\in\cB_n^1\}+\Pr\{B^n\in\cB_n^2\}}\in[1/2:1].$$
Moreover, since
$\{N_k=n\}=\{B^n\in\cB_n\}=\{B^n\in\cB_n^1\}\cup\{B^n\in\cB_n^2\}$ and
$\cB_n^1$ and $\cB_2^n$ are disjoint (by definition),
\begin{multline} \label{eq:pcondn}
  P_{V^n|N_k=n}(v^n) = \rho_n \Pr\{V^n=v^n|B^n\in\cB_n^1\}  \\
  \qquad + 
  (1-\rho_n) \Pr\{V^n=v^n|B^n\in\cB_n^2\}.
\end{multline}
Given the specification of the encoder, we have,
\begin{align}
  &\Pr\{V^n=v^n,B^n\in\cB_n^1\}\nonumber\\
  &\quad = 
  \frac{1}{2^k} \sum_{u^\ast\in\cC_k}
  \Pr\{V^n=v^n,B^n\in\cB_n^1|U^n=u^n\}
  \nonumber\\
  &\quad = 
  \frac{1}{2^k} \sum_{u^\ast\in\cC_k} \sum_{b^n\in\cB_n^1}
  \Pr\{V^n=v^n,B^n=b^n|U^n=u^n\}\nonumber\\
  &\quad = 
  \frac{1}{2^k} \sum_{u^\ast\in\cC_k} \sum_{b^n\in\cB_n^1}
  \ind\{v^n=b^n\oplus u^n\} \Pr\{B^n=b^n\}\nonumber\\
  &\quad\stackrel{(\ast)}{=}
  \frac{\Pr\{B^n\in\cB_n^1\}}{|\cB_n^1|}
  \frac{1}{2^k} \sum_{u^\ast\in\cC_k} \sum_{b^n\in\cB_n^1}
  \ind\{v^n=b^n\oplus u^n\},\nonumber
\end{align}
where $(\ast)$ follows since $\Pr\{B^n=b^n\}$ only depends on
$\weight(b^n)$ and all $b^n\in\cB_1^n$ have the same Hamming weight.  As a
consequence, 
\begin{align}
  \Pr\{V^n=v^n|B^n\in\cB_n^1\} 
  & =
  \frac{1}{|\cB_n^1| 2^k}
  \sum_{u^\ast\in\cC_k} \ind\{u^n\oplus v^n\in\cB_n^1\} \nonumber\\
  &= \frac{1}{|\cB_n^1|2^k} \typen_k(v^n|\cB_n^1) \label{eq:pcondb1}
\end{align}
where for any $\cA_n\subseteq\{0,1\}^n$, we have defined
\begin{equation}
  \typen_k(v^n|\cA_n) \triangleq
  |\{w\in\{0,1\}^k: u^n(w^k) \oplus v^n \in \cA_n\}|.
\end{equation}
We, similarly, have
\begin{equation} \label{eq:pcondb2}
  \Pr\{V^n=v^n|B^n\in\cB_n^2\} =
  \frac{1}{|\cB_n^2| 2^k} \typen_k(v^n|\cB_n^2).
\end{equation}
At this point, we are ready to bound the output divergence using the same
method as in \cite{bastani:2015,bastani:2016}.
Since $P_V^n(v^n)=2^{-n}$, combining \eqref{eq:pcondb1} and
\eqref{eq:pcondb2}, together with the fact that
$|\cB_n^1|=|\cB_n^2|=\frac12|\cB_n|$
in \eqref{eq:pcondn}, we get
\begin{multline}\label{eq:ldef}
  L(v^n) \triangleq 
  \frac{P_{V^n|N_k=n}(v^n)}{P_V^n(v^n)} \\= 
  \frac{2^{n-k}}{\frac12|\cB_n|} \bigl[ \rho_n  \typen_k(v^n|\cB_n^1) +
  (1-\rho_n) \typen_k(v^n|\cB_n^2)\bigr].
\end{multline}
We also recall that
\begin{equation}\label{eq:divl}
  D(P_{V^n|N_k=n}\|P_V^n) = \sum_{v^n} P_V^n(v^n) L(v^n) \log L(v^n).
\end{equation}
Assume the code shared by the encoding functions $(\enc^{(k)}_n,n\in\NN)$
is sampled from i.i.d.\ random coding ensemble, namely, each codeword
$U^\infty(w^k)$ is an infinite i.i.d.\ sequence of binary digits where each
symbol is equally likely to take either value and the codewords are
independent of each other.  In this case,
$\{\typen_k(v^n|\cB_n^1),\typen_k(v^n|\cB_n^2)\}$ forms a multinomial
collection with cluster size $2^k$ and (equal) success probabilities
$2^{-n} \frac12|\cB_n|$.  Thus, it can immediately be verified that
$\overline{L(v^n)}=1$ (where $\overline{A}$ denotes the ensemble average of
$A$).

As shown in \cite{bastani:2015}, since $\overline{L(v^n)}=1$, and
$L(v^n)\le2^n$,
\begin{equation}\label{eq:lloglmin}
  \overline{L(v^n)\log L(v^n)} 
  \le \min\Bigl\{n, \frac{1}{\ln(2)}\overline{(L(v^n)-1)^2}\Bigr\}.
\end{equation}
Since $\typen_k(v^n|\cB_n^1)$ and $\typen_k(v^n|\cB_n^2)$
are negatively correlated, 
\begin{equation} \label{eq:varbound}
  \overline{(L(v^n)-1)^2} \le 2(\rho_n^2+(1-\rho_n)^2)
  \frac{2^{-(k-n)}}{|\cB_n|} \le 2 \frac{2^{-(k-n)}}{|\cB_n|}
\end{equation}
Using \eqref{eq:varbound} in \eqref{eq:lloglmin} and the linearity of the
expectation together with \eqref{eq:divl} we conclude that
\begin{equation}
  \overline{D(P_{V^n|N_k=n}\|P_V^n)} \le
  \min\Bigl\{n, \frac{2}{\ln(2)} \frac{2^{-(k-n)}}{|\cB_n|}\Bigr\}. 
  \label{eq:divnbound}
\end{equation}

Since $\Pr\{N_k=n\}=\Pr\{B^n\in\cB_n^1\}+\Pr\{B^n\in\cB_n^2\}$ and
$\Pr\{B^n\in\cB_n^1\}\ge\Pr\{B^n\in\cB_n^2\}$ (cf.~\eqref{eq:probb}), 
\begin{multline}\label{eq:probnbound}
  \Pr\{N_k=n\} \le 2 \Pr\{B^n\in\cB_n^1\}\\ = 2 |\cB_n|p^{\lfloor nq^*_k(n)
  \rfloor}(1-p)^{n-\lfloor n q^*_k(n) \rfloor}.
\end{multline}

Multiplying the right-hand-sides of \eqref{eq:divnbound} and
\eqref{eq:probnbound} we get
\begin{align} 
  &\overline{D(P_{V^n|N_k=n}\|P_V^n)} \Pr\{N_k=n\} \nonumber\\
  &\quad\le \kappa_1 \min\Bigl\{%
    n |\cB_n| p^{n q_k^*(n)} (1-p)^{n(1-q_k^*(n))},\nonumber\\
    &\quad\qquad\phantom{\min} 
    2^{-(k-n)} p^{n q_k^*(n)} (1-p)^{n(1-q_k^*(n))}%
  \Bigr\}\nonumber\\
  &\quad\aeq  
  \kappa_1 2^{n f_2(q^*_k(n)\|p)}
  \min\Bigl\{n|\cB_n|2^{-n},2^{-k}\Bigr\}\nonumber\\
  &\quad\ble
  \kappa_1 2^{n f_2(q^*_k(n)\|p)}
  \min\Bigl\{n 2^{-n c_2(q_k^*(n))},2^{-k}\Bigr\}\nonumber\\ 
  &\quad\ceq
  \kappa_1 2^{n f_2(q^*_k(n)\|p)}
  \min\Bigl\{n 2^{-k/\alpha},2^{-k}\Bigr\}\nonumber\\ 
  &\quad\le 
  \kappa_1 2^{-k \max\{1,1/\alpha\}} n  2^{n f_2(q^*_k(n)\|p)}
  \label{eq:dtimesp}
\end{align}
where $\kappa_1 = \frac{4}{\ln(2)} \frac{p}{1-p}$,  in (a) we have defined
\begin{equation} \label{eq:f2def}
  f_2(q\|p) \triangleq 1 + q \log (p) + (1-q) \log(1-p),
\end{equation}
(b) follows since $\cB_n$ is a subset of all binary sequences of length $n$
and Hamming weight $nq_k^*(n)$, and  (c) by replacing $n=\frac{k}{\alpha
c_2(q_k^*(n))}$.   Plugging \eqref{eq:dtimesp} into \eqref{eq:divdef}
(noting that the stopping rule is 
independent of the choice of the code) we get
\begin{equation}
  \overline{\DD_k} \le \kappa_1 2^{-k\max\{1,1/\alpha\}}  
\sum_{n\in\cN_k} n  2^{n f_2(q_k^*(n)\|p)}. \label{eq:dbound1}
\end{equation}
Let
\begin{equation}
  \tau_k \triangleq \frac{\log(1-1/k)-[1+\log(1-p)]}{\log(p) - \log(1-p)},
\end{equation}
so that $f_2(\tau_k\|p) = \log(1-1/k)$.  It is easy to verify that
$\tau_k$ is a decreasing sequence and $\tau_k \in (p:1/2)$. Let
$\cN_k^1\triangleq\{n\in\cN_k:q_k^*(n) < \tau_k\}$ and
$\cN_k^2\triangleq\{n\in\cN_k:q_k^*(n) \ge \tau_k\}$, and split the
summation in the right-hand-side of \eqref{eq:dbound1} as 
\begin{multline}
  \sum_{n\in\cN_k} n 2^{n f (q_k^*(n)\|p)} 
  \\ =
  \sum_{n\in\cN_k^1} n 2^{n f_2(q_k^*(n)\|p)}  +
  \sum_{n\in\cN_k^2} n 2^{n f_2(q_k^*(n)\|p)}
  \label{eq:dboundsplit}
\end{multline}
Since $q_k^*(n)$ is increasing in $n$,
\begin{align}
  \sum_{n\in\cN_k^2} n 2^{n f_2(q_k^*\|p)} & \ale 
  \sum_{n \ge \frac{k}{\alpha c_2(\tau_k)}} n 2^{n f_2(q_k^*(n)\|p)} 
  \nonumber \\
  & \ble \sum_{n \ge \frac{k}{\alpha c_2(\tau_k)}} n 2^{n f_2(\tau_k\|p)}
  \le 
  \sum_{n=0}^{\infty} n  2^{n f_2(\tau_k\|p)} \nonumber\\
  & \ceq
  \frac{2^{-f_2(\tau_k\|p)}}{\bigl(2^{-f_2(\tau_k\|p)}-1\bigr)^2} 
  \deq k(k-1) \label{eq:dboundsum2}
\end{align}
where (a) follows since since we included $n\not\in\cN_k$ in the sum as
well, (b) since $f_2(q\|p)$ is decreasing in $q$,  (c) since $f_2(\tau_k\|p)
< 0$ (thus the sum converges) and (d) by replacing
$f_2(\tau_k\|p)=\log(1-1/k)$.

The first summation in \eqref{eq:dboundsplit} has (strictly) less than
$$\frac{k}{\alpha}\frac{1}{c_2(\tau_k)} < 
\frac{k}{\alpha}\frac{1}{c_2(\tau_\infty)}\triangleq \kappa_2(k)$$ terms where
$$\tau_\infty \triangleq \lim_{k\to\infty}\tau_k =
\frac{\log(1-p)+1}{\log(1-p)-\log(p)}.$$  Replacing $n =
\frac{k}{\alpha c_2(q^*_k(n))}$, we see that each term in the first
summation of \eqref{eq:dboundsplit} is upper-bounded as
\begin{equation}
  n 2^{n f_2(q_k^*(n)\|p)}  \le \kappa_2(k) 
  2^{k \frac{f_2(q_k^*(n)\|p)}{\alpha c_2(q_k^*(n))}} 
  \le \kappa_2(k) 2^{k/\alpha}
\end{equation} 
with equality iff $q_k^*(n) = p$. (This term is included in the
summation since $\tau_k > p$.)  Indeed, the last step follows since
$f_2(q\|p) = c_2(q) - d_2(q\|p)$.
Consequently, 
\begin{equation}
  \sum_{n\in\cN_k^1}
  n 2^{n f_2(q_k^*(n)\|p)} 
  \le
  \kappa_2(k)^2 2^{k/\alpha}
  \label{eq:dboundsum1}
\end{equation}
Combining \eqref{eq:dboundsum2} and \eqref{eq:dboundsum1} (noting that the
right-hand-side of \eqref{eq:dboundsum1} grows faster than that of
\eqref{eq:dboundsum2}) shows that, for large $k$,
\begin{equation} \label{eq:dbound2}
  \overline{\DD_k} \le 
  2 \kappa_1 \kappa_2(k) 2^{-k [\max\{1,\frac{1}{\alpha}\}-\frac{1}{\alpha}]} = 
  \kappa_3(k) 2^{-k \frac{[\alpha-1]^+}{\alpha}},
\end{equation}
where we have defined $\kappa_3(k)\triangleq 2\kappa_1\kappa_2(k)$.
Therefore, for at least half of the codes,
\begin{equation} \label{eq:dcode}
  \DD_k \le 2 \overline{\DD_k} \le 2\kappa_3(k)
  2^{-k\frac{[\alpha-1]^+}{\alpha}}.
\end{equation}
Since $\lim_{k\to\infty}\frac1k\log(\kappa_3(k))=0$, by picking any such
good code for each $k$ we will have a sequence of codes for which
\begin{equation}
  \liminf_{k\to\infty}\frac{-\log \DD_k}{k} \ge
  \frac{[\alpha-1]^+}{\alpha}.
  \label{eq:expk}
\end{equation}
Equations \eqref{eq:stop:inf} and \eqref{eq:expk} imply
\begin{equation*}
  \liminf_{k\to\infty}\frac{-\log \DD_k}{\E[N_k]}  = 
  \liminf_{k\to\infty}\frac{-\log \DD_k}{k}\frac{k}{\E[N_k]}
  \ge [\alpha-1]^+ c_2(p).
\end{equation*}
Setting $\alpha=R/c_2(p)$ proves Theorem~\ref{thm:feedbackexp}. \hfill\IEEEQED
\section{Conclusion and Discussion}
We studied the problem of channel resolvability in presence of feedback.
We showed that, while feedback does not decrease the channel resolution, in
presence of causal feedback higher \emph{resolvability exponents} compared
to the existing block resolvability codes of
\cite{hayashi:2006,hou:2013,han:2014,hayashi:2015arxiv,bastani:2015,bastani:2016}
are achievable.

Our results are the analogue of establishing the achievability of the error
exponent $[I(U;V)-R]^+$  in presence
of feedback (cf.\ \cite[Section~2.1]{tchamkerten:thesis}) for 
channel coding. (Burnashev's exponent \cite{burnashev:1976} is also
a straight line but with a steeper slope.)
However, since, to the best of our knowledge, no non-trivial upper bounds
on the highest achievable resolvability exponent at a specific rate $R$
(i.e., an equivalent of sphere-packing exponent for channel coding)  is
known, it is unclear whether the improvement we demonstrated in this work
is exclusively due to the presence of feedback or there might exist a
resolvability scheme that achieves the straight-line exponent of
\eqref{eq:feedbackexp} without the need for feedback.  Nevertheless, the
results of \cite{bastani:2016} show that an average i.i.d.\ random code
cannot achieve a better resolvability exponent than \eqref{eq:blockexp}.
Thus, at least for the i.i.d.\ random coding ensemble,
the gains in the exponent are due to the presence of feedback.  

Moreover, for the channel coding problem, Dobrushin \cite{dobrushin:1963}
and Haroutunian \cite{haroutunian:1977} upper-bounded the best
attainable error exponent in presence of feedback using block codes
(This upper bound equals the sphere-packing exponent for symmetric channels
  \cite{dobrushin:1963} but is larger than that, for asymmetric ones
\cite[Exercise~10.36]{csiszar:it}.)  Those results imply that employing
variable-length error correcting codes is necessary to achieve the higher
exponents of \cite{burnashev:1976}.  Another important subject for future
research is to study the achievable resolvability exponents using block
resolvability codes in presence of feedback.
\section*{Acknowledgment}
This work was supported by the Swiss National Science Foundation under
grant number 200020\_146832.
\appendix
\begin{lemma} \label{lem:stopvar}
  Let $(\xi_n,~n\in\NN)$ be i.i.d.\ zero-mean random variables and
  $$S_n\triangleq\sum_{i=1}^{n} \xi_n,\qquad n\in\NN.$$
  Then the process $(S_n,~n\in\NN)$ is a martingale with respect to the
  natural filtering $\bigl(\cF_n=\sigma(\xi_1,\dots,\xi_n),~n\in\NN\bigr)$
  and, if $N$ is a stopping time,
  \begin{equation}
    \E\Bigl[\frac{S_N^2}{N}\Bigr] \le \var(\xi_1) \E[1+\ln(N)].
    \label{eq:stopvar}
  \end{equation}
\end{lemma}
\begin{IEEEproof}
  That $(S_n,~n\in\NN)$ is a martingale is trivial.  We shall only prove
  \eqref{eq:stopvar}.  Let
  \begin{equation*}
    N_{m} \triangleq \min\{N,m\}, \qquad \forall m\in\NN.
  \end{equation*}
  It is clear that $\forall m\in\NN$, $N_{m} \in
  \indexset{1:m}$ almost surely and $N_m$ is a stopping time.  The latter
  can be verified by noting that
  \begin{equation} \label{eq:stopvar:cases}
    \{N_m=n\} =
    \begin{cases}
      \{N=n\} & \text{if $n < m$},\\
      \{N\ge m\} & \text{if $n=m$}.
    \end{cases}
  \end{equation}
  Thus for $n<m$, $\{N_m=n\}=\{N=n\}\in\cF_n$ by the hypothesis that $N$
  is a stopping time, and for $n=m$,
  \begin{align}
    \{N_m=m\} = \{N\ge m\}
    = \bigcap_{j=1}^{m-1} \{N \ne j\} \in \cF_{m-1},
    \label{eq:stopvar:indep}
  \end{align}
  and $\cF_{m-1}\subseteq\cF_m$ (hence $\{N_m=n\}\in\cF_m$).
  Finally $N_1=1$ almost surely, hence,
  \begin{equation}
    \E\biggl[\frac{S_{N_1}^2}{N_1}\biggr] = \var(\xi_1).
    \label{eq:stopvar:last}
  \end{equation}
  We now have
  \begin{align}
    & \E\biggl[\frac{S_{N_m}^2}{N_m}\biggr]-
    \E\biggl[\frac{S_{N_{m-1}}^2}{N_{m-1}}\biggr]
    =
    \E\biggl[\biggl(\frac{S_m^2}{m} -
    \frac{S_{m-1}^2}{m-1}\biggr)\ind\{N\ge m\}\biggr]
    \nonumber\\
    &\quad=
    \E\biggl[\frac{(m-1) \bigl(\xi_{m}^2+2\xi_{m}S_{m-1}\bigr)-S_{m-1}^2} {(m-1)m}
    \ind\{N\ge m\}\biggr]\nonumber\\
    &\quad\le
    \frac{1}{m}\bigl(\E[\xi^2_{m}\ind\{N\ge m\}]
      +2\E[\xi_{m} S_{m-1}\ind\{N\ge m\}]
    \bigr)\nonumber\\
    &\quad \stackrel{(\ast)}{=} \frac{1}{m} \var(\xi_{m})
    \Pr\{N\ge m\}.
    \label{eq:stopvar:singlestep}
  \end{align}
  In the above $(\ast)$ follows since, as shown in \eqref{eq:stopvar:indep},
  $\{N\ge m\}\in\cF_{m-1}$ thus $\ind\{N\ge m\}$ is independent of
  $\xi_{m}$.

  Using \eqref{eq:stopvar:singlestep} repeatedly together with the fact
  that $\forall n\in\NN\colon \var(\xi_n)=\var(\xi_1)$, we get
  \begin{align}
    \E\biggl[\frac{S_{N_m}^2}{N_m}\biggr] &\le
    \E\biggl[\frac{S_{N_1}^2}{N_1}\biggr] +
    \var(\xi_1)\sum_{\ell=2}^m\frac{\Pr\{N\ge \ell\}}{\ell}
    \nonumber\\
    & \stackrel{(\ast)}{=}
    \var(\xi_1)\sum_{\ell=1}^{m}\frac{\Pr\{N\ge \ell\}}{\ell}\nonumber\\
    & \le \var(\xi_1) \sum_{\ell\ge1} \frac{\Pr\{N\ge\ell\}}{\ell}.
    \label{eq:stopvar:finalsum}
  \end{align}
  where $(\ast)$ follows from \eqref{eq:stopvar:last} and the fact that
  $N\ge1$ almost surely.  We finally have
  \begin{align}
    &\sum_{\ell\ge1} \frac{\Pr\{N\ge \ell\}}{\ell}  =
    \sum_{n\ge1} \Pr\{N=n\} \sum_{\ell=1}^{n} \frac1\ell
    \nonumber\\
    &\quad \le
    \sum_{n\ge1} \Pr\{N=n\} (1+\ln(n)) = \E[1+\ln(N)].
  \end{align}
  Using the above in \eqref{eq:stopvar:finalsum} yields
  \begin{equation}
    \E\biggl[\frac{S_{N_m}^2}{N_m}\biggr] \le \E[1+\ln(N)], \qquad 
    \forall m\in\NN.
    \label{eq:stopvar:finite}
  \end{equation}
  Now, since $\lim_{m\to\infty} N_n = N$ with probability $1$ 
  \begin{multline}
    \E\Bigl[\frac{S_N^2}{N}\Bigr] = 
    \E\biggl[\lim_{m\to\infty} \frac{S_{N_m}^2}{N_m}\biggr]  =
    \E\biggl[\liminf_{m\to\infty} \frac{S_{N_m}^2}{N_m}\biggr] \\
    \ale \liminf_{m\to\infty}
    \E\biggl[\frac{S_{N_m}^2}{N_m}\biggr] 
    \ble \E[1+\ln(N)].
  \end{multline}
  where in the above (a) follows from Fatou's lemma (applied to the
    sequence of non-negative random variables $\frac{S_{N_m}^2}{N_m}$,
  $m\in\NN$) and (b) from \eqref{eq:stopvar:finite}.
\end{IEEEproof}



\bibliographystyle{IEEEtran}
\bibliography{IEEEabrv,confabrv,../references}

\end{document}